\documentclass[11pt,a4paper]{article}
\usepackage{jcappub}
\usepackage{xspace}
\usepackage{color}
\usepackage{graphicx}
\usepackage{booktabs}
\title{The small scale power asymmetry\\ in the cosmic microwave background}

\author{Samuel Flender,}
\author{Shaun Hotchkiss}
\affiliation{Department of Physics, University of Helsinki and Helsinki Institute of Physics, \\
P.O. Box 64, FIN-00014, University of Helsinki, Finland}

\emailAdd{samuel.flender@helsinki.fi}

\abstract{
We investigate the hemispherical power asymmetry in the cosmic microwave background on small angular scales. We find an anomalously high asymmetry in the multipole range $\ell=601-2048$, with a naive statistical significance of 6.5$\sigma$. However, we show that this extreme anomaly is simply a coincidence of three other effects, relativistic power modulation, edge effects from the mask applied, and inter-scale correlations. After correcting for all of these effects, the significance level drops to $\sim1\sigma$, i.e., there is no anomalous intrinsic asymmetry in the small angular scales. Using this null result, we derive a constraint on a potential dipolar modulation amplitude, $A(k)<0.0045$ on the $\sim10\,\rm Mpc$-scale, at 95\% C.L. This new constraint must be satisfied by any theoretical model attempting to explain the hemispherical asymmetry at large angular scales.
}

\arxivnumber{1307.6069}

\begin{document}
\maketitle
\flushbottom

\newcommand{\muk}{\,\rm{{\mu}K}}
\newcommand{\Mpc}{\,\rm{Mpc}}
\newcommand{\mpc}{\,h^{-1}\,\rm{Mpc}}
\newcommand{\dg}{^{\circ}}
\newcommand{\lcdm}{$\Lambda$CDM\xspace}
\newcommand{\fnl}{f_{\mathrm{NL}}}
\newcommand{\beq}{\begin{equation}}
\newcommand{\eeq}{\end{equation}}
\newcommand{\smica}{\texttt{SMICA}\xspace}
\newcommand{\sevem}{\texttt{SEVEM}\xspace}
\newcommand{\nilc}{\texttt{NILC}\xspace}
\newcommand{\comrul}{\texttt{C-R}\xspace}
\newcommand{\healpix}{\texttt{HEALPix}\xspace}
\newcommand{\apr}{\emph{a priori}\xspace}
\newcommand{\apo}{\emph{a posteriori}\xspace}
\newcommand{\dbll}{\bar\Delta_{\rm ll}}
\newcommand{\dbhl}{\bar\Delta_{\rm hl}}
\newcommand{\dbrel}{\bar\Delta_{\rm rel}}
\newcommand{\dbisc}{\bar\Delta_{\rm isc}}

%---------------Introduction------------------%

\section{Introduction}
\label{section:intro}

The statistical properties of the anisotropies in the temperature of the cosmic microwave background (CMB) possess a wealth of cosmological information. There are many potential methods to extract this information, each of which will provide statistical information about a particular aspect of cosmology. If the universe is statistically homogeneous and isotropic, and the primordial temperature anisotropies follow a Gaussian distribution, then the angular power spectrum is sufficient to describe all the statistical properties of these anisotropies. Therefore, in such a universe, a measurement of the angular power spectrum would also provide all the cosmological information available within the CMB temperature anisotropies.

The angular power spectrum of the measured CMB temperature anisotropies fits its theoretical prediction from the statistically Gaussian, homogeneous and isotropic, $\Lambda$ cold dark matter (\lcdm) standard model, remarkably well \cite{Ade:2013zuv}. From the angular power spectrum alone, no strong evidence is found for the existence of, e.g., additional matter or radiation species, isocurvature in the primordial fluctuations, or a primordial spectrum of curvature fluctuations that is not described by a power-law. However, some anomalies have been found using other statistical observables, i.e. features in the observed sky that have a low probability of occurring in the \lcdm model. These anomalies are predominantly seen on the largest angular scales. An overview of anomalies seen in the WMAP data is presented in \cite{Bennett:2010jb, Copi:2010na}. Most of the anomalies seen in the WMAP data have  been confirmed with the new Planck data \cite{Ade:2013nlj}.

One such anomaly is the `hemispherical power asymmetry': Eriksen et al \cite{Eriksen:2003db} and Hansen et al \cite{Hansen:2004vq} found that the angular power spectrum of the CMB temperature anisotropies, from the 1-year WMAP data, appears itself to be anisotropic. In particular it was found that, in the low multipole range ${\ell}=2-40$, the hemisphere centred at Galactic coordinates $(l,b)=(237\dg,-20\dg)$ has significantly more power than the one opposite on the sky. A possible connection between this hemispherical asymmetry and an alignment between the quadrupole and octopole was pointed out in \cite{schwarz:2004}. It has also been shown that the power asymmetry can be described in terms of a dipolar modulation \cite{Gordon:2005ai, Hoftuft:2009rq}. In \cite{Paci:2013gs} it was shown that there is no similar asymmetry in the WMAP polarisation data. In \cite{Eriksen:2004iu} the hemispherical asymmetry was confirmed by analysing large angular scale $N$-point correlation functions, which act in real (pixel) space instead of harmonic space. Following this approach, the Planck collaboration confirmed the existence of a hemispherical asymmetry in the multipole range ${\ell}=2-40$ \cite{Ade:2013nlj}. The reported $p$-value (i.e. the probability to find an asymmetry at least as strong as the one measured in our CMB) is $\lesssim 0.01$.

Later it was found that this hemispherical power asymmetry persists up to much smaller scales \cite{Hansen:2008ym}. In particular, for the whole multipole range ${\ell}=2-600$, significantly more power was found in a disc centered at $(l,b)=(226\dg,-27\dg)$ than in one centred on the opposite point of the sky. The authors report a $p$-value of $0.004$. In the recent Planck results it was suggested that the power asymmetry persists in an even larger multipole range, ${\ell}=2-1500$, in the direction $(l,b)=(224\dg,0\dg)$ \cite{Ade:2013nlj}. The authors calculate the localised power in discs and introduce a measure for the asymmetry which is based on the clustering of the directions of maximal asymmetry in different multipole bins. The reported $p$-value is $\lesssim 0.008$.

Judging these detections of a hemispherical asymmetry is difficult. The reported $p$-values are small enough to be interesting, but none yet are sufficiently small to rule out the possibility of a statistical fluke. Planck has now measured the temperature of the CMB in $\sim 10^7$ pixels. The number of different independent statistical methods that can be used to analyse Planck's maps of the temperature anisotropies is therefore immense. It is \emph{expected} that if enough statistical tests are applied, some of them will yield results that would be judged as anomalous, even by the correct cosmological model. Therefore, we should not be surprised if our own universe happens to possess a few anomalous aspects when analysed using sufficiently many different statistical measures. Conversely, however, it is also expected that if the \lcdm model is not entirely correct, evidence for the new physics beyond \lcdm would first show up as an anomaly. Therefore, these anomalies are worth consideration, however care must be taken in their interpretation.

How can we conclusively determine the true origin of these anomalies? The obvious answer is to simply take more data. And, even for the CMB, this is in principle possible, if we wait long enough. Within $10^2-10^3$ years, well within the time-frame of recorded human history, the CMB arriving at Earth will be from sufficiently farther away that small changes in the anisotropies will be measurable \cite{Moss:2007bu,Lange:2007tx}. If the large scale anomalies remain equally significant, this will be evidence in favour of new physics, and if not, this will be evidence in favour of a statistical fluke.

In the near future, however, the situation is more bleak. We only have one sky, and the large scale fluctuations in its temperature have already been measured, although some systematic differences between the CMB maps from Planck and WMAP have been pointed out \cite{Frejsel:2013nfa}, indicating that even on the largest scales some improvements might still be possible. The best method is to construct models that are able to generate some or all of the anomalies. Then, a full Bayesian analysis can be performed comparing the new class of models to \lcdm (see, e.g., \cite{Hoftuft:2009rq}). However, because the anomalies are only anomalous in \lcdm at the $~\%$-level, this will still remain inconclusive unless a new model can postdict precisely the observed anomalies with few (or no) extra parameters. Furthermore, if a plausible model predicts other deviations from \lcdm, these can be searched for to provide additional evidence for or against the new model. Another possible method, the one we apply here, is to search for physical effects very similar to the observed anomalies. For example, if one observable provides a preferred direction, it is worthwhile examining whether other observables also prefer that direction. Or, if there is an alignment in some angular multipoles, it is worth looking for alignments in other multipoles too.

For the hemispherical power asymmetry, theoretical models capable of explaining the origin of the anomaly have arisen within the context of inflation (see, e.g. \cite{Gordon:2006ag, Erickcek:2008jp, Lyth:2013vha, Liu:2013kea, Mazumdar:2013yta, Namjoo:2013fka, Chang:2013vla}). Depending on parameters, these models would also generate a hemispherical power asymmetry at small scales. Planck has now measured the temperature of the CMB on angular scales corresponding to ${\ell}\gtrsim 2000$. If it were to be found that the hemispherical asymmetry persisted down to the smallest scales, with an amplitude that was anomalous in the \lcdm model, this would provide strong evidence for these models and the interpretation that the large scale asymmetry is a sign of new physics. If not, then it does not immediately determine that the large scale asymmetry is a fluke; however it does provide a constraint that must be simultaneously satisfied by any model attempting to explain the large scale asymmetry.

In this work we investigate the small angular scales as a new source of information about the hemispherical asymmetry. Using a simple asymmetry measure, we confirm a large scale hemispherical asymmetry  in the `low multipoles' ${\ell}=2-600$, with a $p$-value $\sim 0.01$. Applying the exact same approach to the `high multipoles' ${\ell}=601-2048$, we initially find an extreme asymmetry with an estimated statistical significance as large as $6.5\sigma$ (corresponding, in a Gaussian distribution, to a $p$-value of $10^{-10}$!). However, after further investigation, we find that this anomalous signal is caused by a coincidence of three effects: relativistic power modulation due to the Galaxy's  motion relative to the CMB rest frame, edge effects due the mask used, and inter-scale correlations due to incomplete sky coverage. After taking into account all of these effects, the significance drops to the $\sim1\sigma$ level - i.e. the small scales do not increase the anomaly's significance.

We conclude that any theoretical model that produces asymmetry on large angular scales must not also produce strong \emph{intrinsic} asymmetry on small angular scales. In particular, with our measure we restrict the small scale asymmetry in the multipole range ${\ell}=601-2048$ to be smaller than $1.53\%$ at $95\%$ C.L. We show that this can be translated into a constraint on the dipolar modulation amplitude, $A(k)<0.0045$ on the $\sim10\,\rm Mpc$-scale, at 95\% C.L.

%----------------------------------%
\section{Methodology}
\label{section:method}

%%%%Fig%%%%
\begin{figure}
\begin{center}
\includegraphics[scale=0.27]{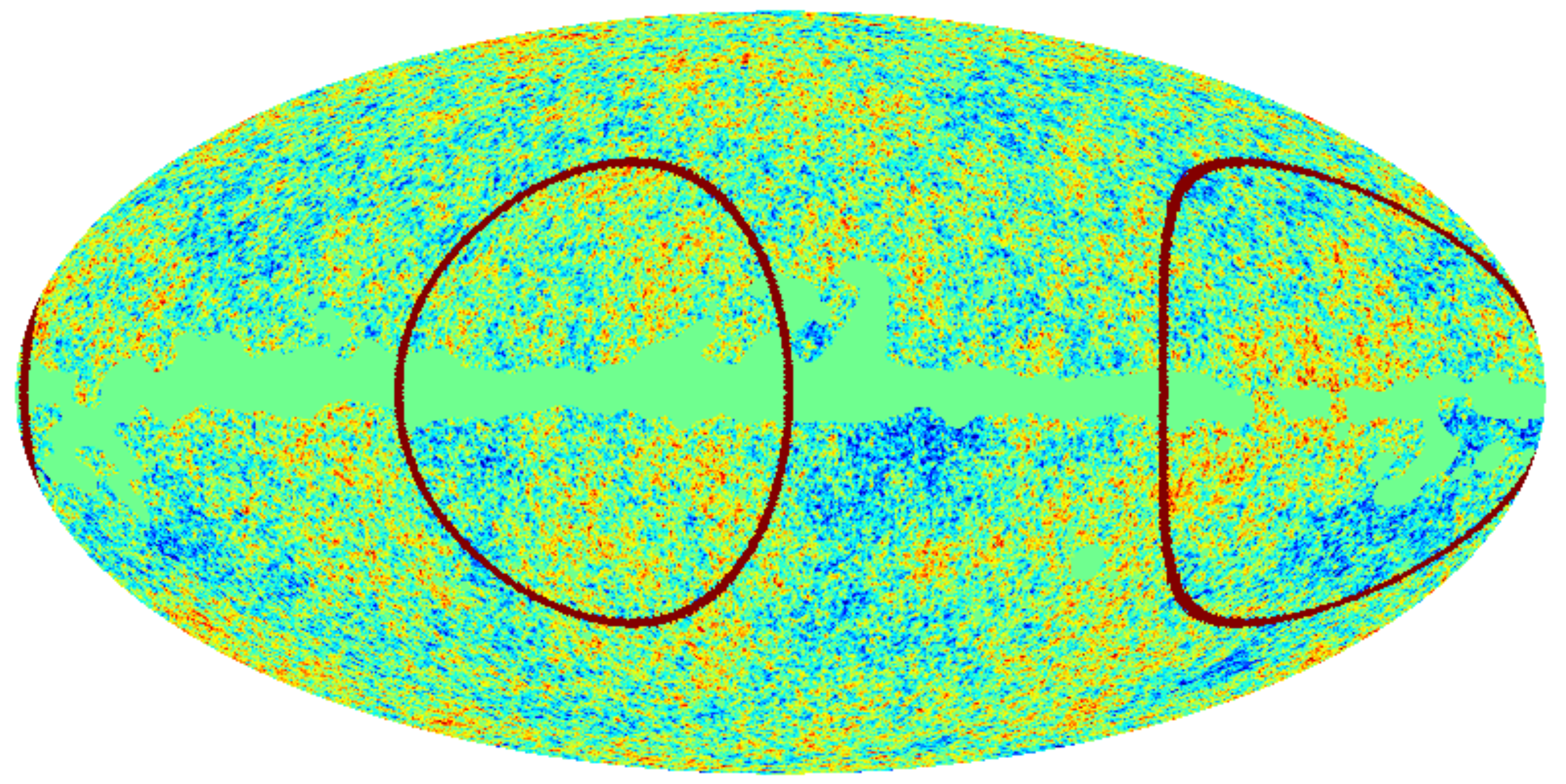}
\includegraphics[scale=0.27]{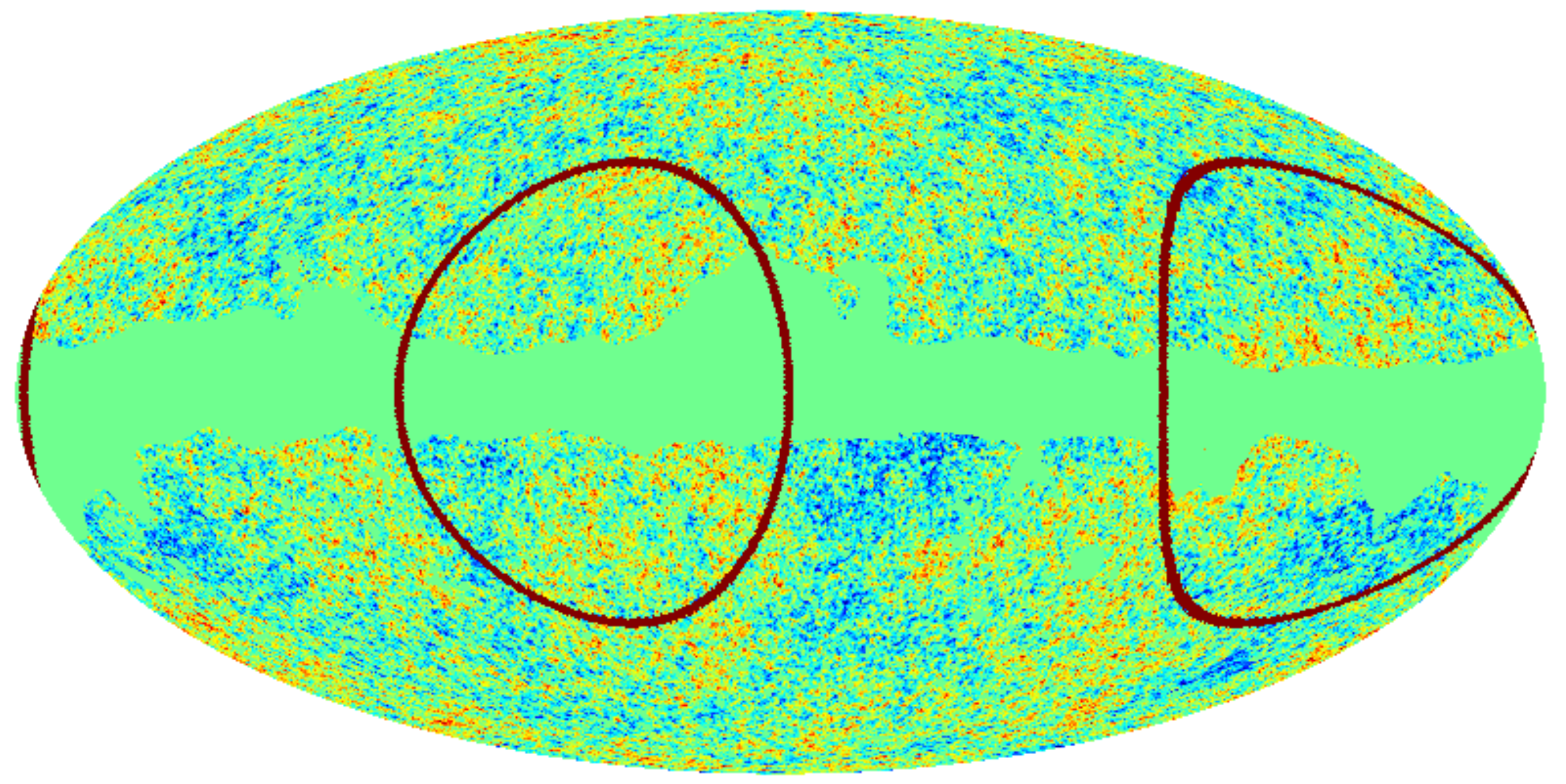}
\caption{\label{figure:locspots_smica} \smica map of the CMB with \smica mask (left) and M74 mask (right) applied. The red circles indicate the locations of the discs along which we calculate the local power asymmetry of the map: $(l,b)=(224\dg,0\dg)$ (right circle) and the opposite direction $(l,b)=(44\dg,0\dg)$ (left circle).}
\end{center}
\end{figure}
%%%%%%%%%%

In order to study the hemispherical power asymmetry, we adopt the method proposed by Hansen et al \cite{Hansen:2008ym} and compare the magnitude of the power spectrum of the temperature anisotropies within two discs of a given diameter at opposite positions of the sky. In their approach, \cite{Hansen:2008ym} vary the disc diameter from $22.5\dg$ to $180\dg$. Here we choose a fixed diameter of $90\dg$, which is a compromise between the amount of data allowed into a disc and computational effort. At no point in our analysis  do we vary this radius. Our motivation for not doing this is to minimise the risk of \apo statistics. It is possible that a more anomalous signal could be found using a different disc radius; however it would then be necessary to quantify the statistical likelihood that an anomalous signal could arise at \emph{any} of the examined radii, rather than just at the one radius where it happened to be maximum in our universe.

We choose the location of the disc to be the direction of maximal asymmetry over the whole multipole range ${\ell}=2-1500$ found by Planck in \cite{Ade:2013nlj}, $(l,b)=(224\dg,0\dg)$. Thus, if the results in \cite{Ade:2013nlj} are correct, we do expect the most extreme hemispherical asymmetry along that line of sight, in both the low multipole range and the high multipole range. One result of this choice is that, inevitably, some form of \apo statistics will leak into our analysis. This would be a problem if we found a large anomaly on the small scales. Given that we in fact find the opposite, this is not a problem. If $(l,b)=(224\dg,0\dg)$ is truly the direction of \emph{maximal} asymmetry, this \emph{strengthens} our final result, which is a null-detection and thus a constraint on how much primordial asymmetry can exist. If, however, another direction is more asymmetric, our constraint still applies for the direction $(l,b)=(224\dg,0\dg)$, where the asymmetry cannot exceed $\sim 1.5 \%$. Any model generating asymmetry still cannot cause the asymmetry to exceed this value, on these scales, at this direction on the sky.

We study the asymmetry in the low multipoles ${\ell}=2-600$ separately from a potential asymmetry in the high multipoles ${\ell}=601-2048$. This way we can identify the new information, regarding this anomaly, which comes solely from the small scales. By keeping the direction $(l,b)=(224\dg,0\dg)$ and the disc diameter fixed throughout our analysis we minimise the risk of \apo statistics, which would erroneously enhance the perceived significance of any anomaly. It is true that the specific direction $(l,b)=(224\dg,0\dg)$ was found by Planck based on information at both the large and small scales; however the significance of the asymmetry along this general direction was recognised before any of the multipoles in the range ${\ell}=601-2048$ had been measured.

For our analysis of the small scales, throughout, we use Planck's \smica map with a resolution of $N_{\rm side}=2048$. To remove the galactic foregrounds and point sources from this map, we use two different masks, the \smica confidence mask, which leaves 89\% of the sky unmasked, and our own mask, `M74', which leaves 74\% of the sky unmasked. We construct M74 by taking the union of the \smica, \sevem and \nilc confidence masks and the \texttt{Commander-Ruler minimal} mask.\footnote{The \smica map and masks are available at: \texttt{http://pla.esac.esa.int/pla/aio/planckProducts.html}} M74 is very similar to the U73 mask used by the Planck collaboration, which is not publicly available. The two masks can be seen in Fig.\,\ref{figure:locspots_smica}. For the calculation of the power spectra we make use of the \healpix software package\footnote{\texttt{http://healpix.jpl.nasa.gov}}.

In order to quantify the hemispherical asymmetry in both the large and small scales, we introduce the following two measures:
\beq
\dbll \equiv \frac{1}{599}\sum_{\ell=2}^{600} \Delta_{\ell} ,
\eeq
which measures the average asymmetry in the low multipoles and
\beq
\dbhl \equiv \frac{1}{1448}\sum_{\ell=601}^{2048} \Delta_{\ell} ,
\eeq
which measures the average asymmetry in the high multipoles. In the above, $\Delta_{\ell} \equiv \frac{C_{\ell}^{+}-C_{\ell}^{-}}{C_{\ell}^{+}}$ is the relative difference between the power $C_{\ell}^+$ in a disc of diameter $90\dg$ centered at $(l,b)=(224\dg,0\dg)$, and the power $C_{\ell}^-$ in a disc in the opposite direction.

We do not claim that this measure gives the largest significance for the anomaly. However, it is a very simple and naturally motivated measure to quantify the overall (average) asymmetry on both the large and small scales. Again, in order to minimise any \apo effects we only use this measure throughout our analysis. As will be seen, it reproduces the large scale asymmetry, making it a useful \apr statistic for examining the new small scale information.

In order to estimate the significance of the asymmetry, we generate 1000 random Gaussian realisations of the CMB, using the \texttt{synfast} facility of the \healpix package, with the input power spectrum obtained by running \texttt{CAMB}\footnote{\texttt{http://camb.info}} with the cosmological parameters from Planck \cite{Ade:2013zuv}. We apply exactly the same methodology to each of the random realisations as we do to the \smica map, i.e. we apply the same masks (\smica mask and M74) and choose the same disc diameter ($90\dg$). Furthermore, we keep the same directions for the discs ($(l,b)=(224\dg,0\dg)$ and opposite) in each simulation in order to account for possible effects caused by the masks along these particular lines of sight. As we will show, the distribution of the values $\dbll$ and $\dbhl$ found in the simulations is, to a good approximation, Gaussian (see e.g. Fig.\,\ref{figure:hist_sm_M74}). It is therefore reasonable to express the significance of the asymmetry in terms of standard deviations from this (approximately) Gaussian distribution. From the significance so obtained we derive the `$p$-value', i.e. the probability to find an asymmetry at least as strong as the one present in our CMB.

By choosing to keep the disc direction fixed in the random realisations we are not able to quantify the probability that the level of asymmetry observed in our universe could have been observed in \emph{any} direction. This is a small problem for the large scale asymmetry, because before it was detected there was no reason to be interested in this particular direction over any other. However, we are primarily interested in the asymmetry on small scales. Here, we know that the direction $(l,b)=(224\dg,0\dg)$ is interesting \apr. The question we are trying to answer in this work is ``given that there is an anomalous large scale asymmetry along the direction $(l,b)=(224\dg,0\dg)$, is the corresponding small scale asymmetry along this line of sight expected or anomalous?'' To properly account for possible mask effects this means we must keep the location of the disc fixed with respect to the mask in each random realisation.

%--------------------------------%
\section{Results}
\label{section:results}

\subsection{Confirmation of the large scale power asymmetry}

%%%%Fig%%%%
\begin{figure}
\begin{center}
\includegraphics[scale=0.8]{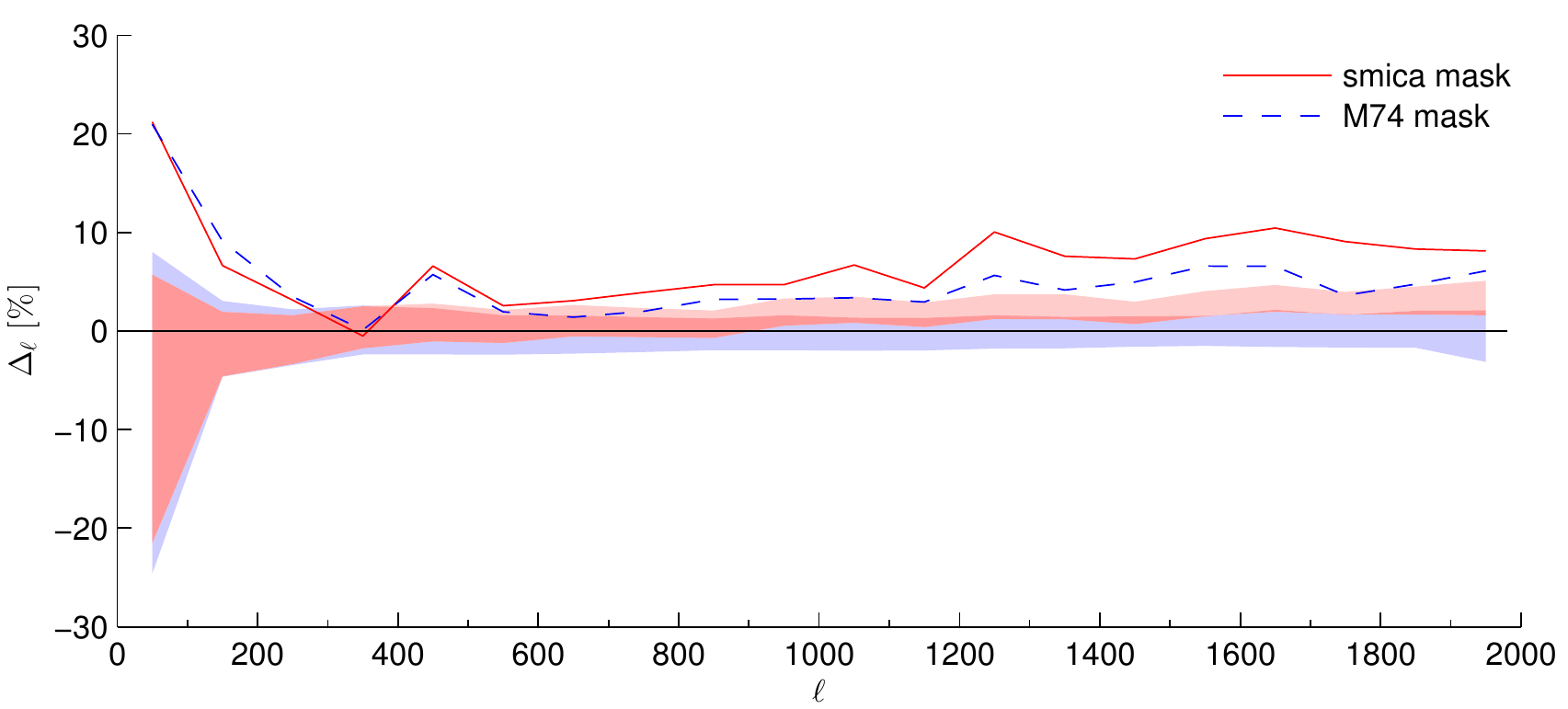}
\caption{\label{figure:Deltal_sm_M74} Binned plot of ${\Delta}_{\ell}$ along $(l,b)=(224\dg,0\dg)$ in the \smica map with \smica mask (red, solid) and M74 mask (blue, dashed) applied. The red and blue bands are the $1\sigma$-regions from the corresponding values found in the simulations. The multipole bin size is $\Delta {\ell} = 100$, starting at ${\ell}=2$. The bias towards positive values in the simulations with the \smica mask applied is caused by edge effects in the \smica mask, as we will discuss in Sec.\,\ref{sec:edgeeffects}. A similar plot can be found in Fig.\,28 in \citep{Ade:2013nlj}.}
\end{center}
\end{figure}
%%%%%%%%%%

We first study the large scale power asymmetry by calculating $\dbll$ in the \smica map. We find  $\dbll=6.62\%\,(3\sigma,\,p\simeq0.003)$ with the \smica mask and $\dbll=6.90\%\,(2.5\sigma,\,p\simeq0.01)$ with the M74 mask applied. The values for different masks are in good agreement, which indicates that there are no obvious mask effects on the large scales.

We thus confirm the existence of an anomalous asymmetry in the CMB at large angular scales ${\ell}=2-600$, which has a statistical significance of $\lesssim 3\sigma$ with our measure. The low-$\ell$ asymmetry is clearly visible in a plot of $\Delta_{\ell}$ as a function of ${\ell}$ (see Fig.\,\ref{figure:Deltal_sm_M74}).

\subsection{Small scale power asymmetry - `naive' approach}

%%%%Fig%%%%
\begin{figure}%[b]
\begin{center}
\includegraphics[scale=0.4]{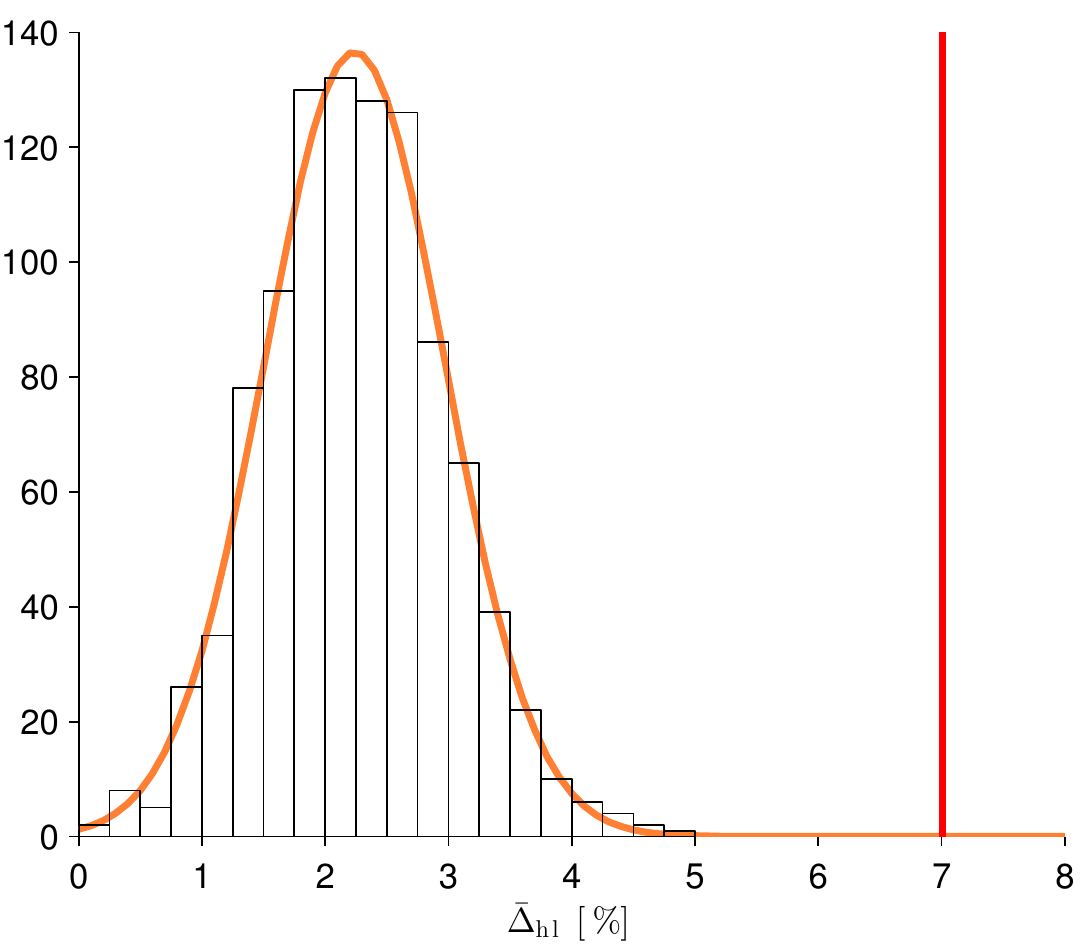}
\includegraphics[scale=0.4]{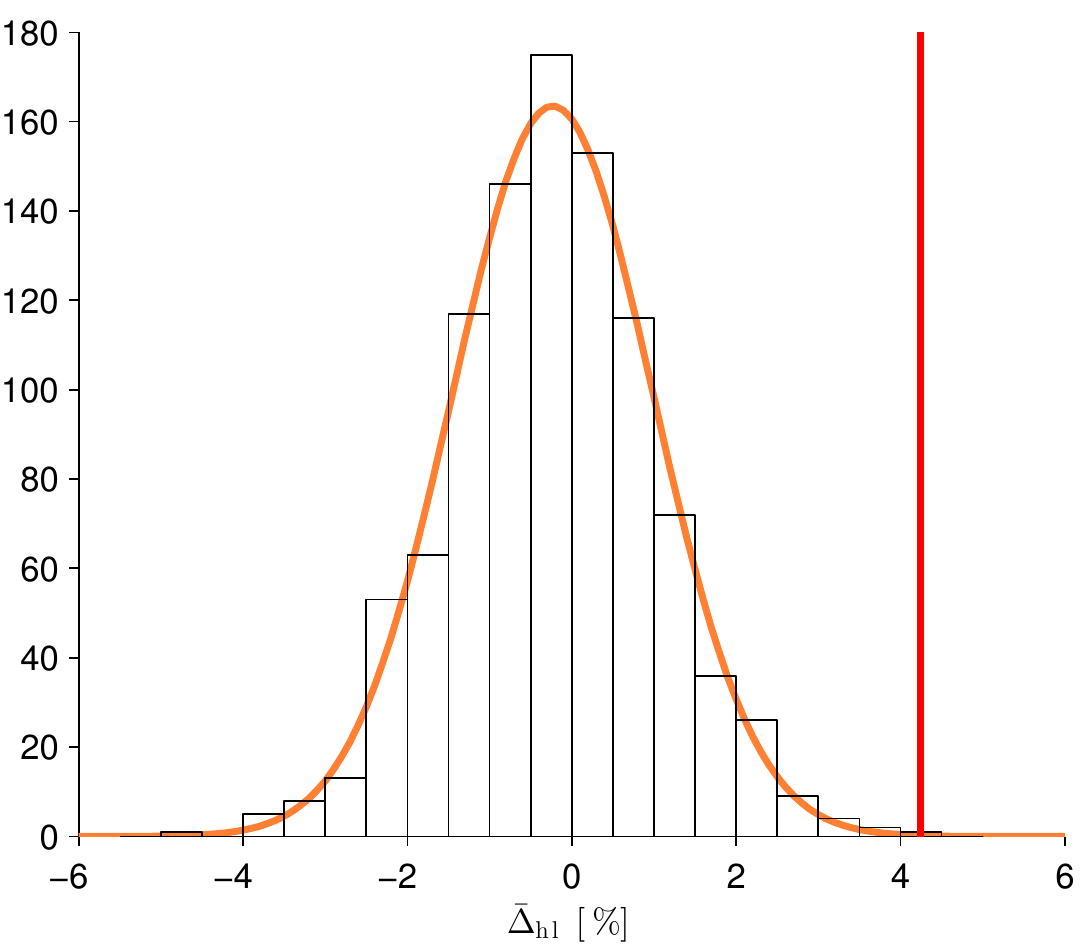}
\caption{\label{figure:hist_sm_M74} Distribution of $\dbhl$ values in 1000 random CMB realisations with \smica mask (left) and M74 mask (right) applied. The orange lines are the corresponding Gaussian probability distribution functions. The red line indicates the value of $\dbhl$ in the \smica map. The non-zero mean of the simulations with the \smica mask applied is a result of the edge effects in this mask, see Sec.\,\ref{sec:edgeeffects}.}
\end{center}
\end{figure}
%%%%%%%%%%

We now keep the direction, $(l,b)=(224\dg,0\dg),$ and the disc diameter, $90\dg$, fixed, but analyse the high multipoles ${\ell}=601-2048$. With the M74 mask we find $\dbhl=4.23\%$. Compared to the simulations this is anomalous with a statistical significance of $3.6\sigma$ ($p\simeq3\cdot10^{-4}$). Using the \smica mask we find a value of $\dbhl=7.01\%$, which is extremely anomalous with an estimated significance of $6.5\sigma$ ($p\sim 10^{-10}$).

Naively, this could be interpreted as a very strong detection of an asymmetry in the high multipoles of the CMB. However, we observe the following oddities in this measurement: Firstly, the value of $\dbhl$ depends on the mask used. This mask dependence is also clearly visible in the behaviour of ${\Delta}_{\ell}$ as a function of ${\ell}$ for multipoles ${\ell}>600$ (see Fig.\,\ref{figure:Deltal_sm_M74}). Secondly, $\Delta_{\ell}$ generally keeps increasing for increasing $\ell$. Finally, the distribution of the $\dbhl$ values obtained from the random realisations does not have a mean of zero when the the \smica mask is applied (see Fig.\,\ref{figure:hist_sm_M74}). This suggests that some of the apparent small scale asymmetry (and, potentially, its significance) could be caused by the masks being used.

In fact there are at least three effects that will contaminate a measurement of $\dbhl$ using Planck's maps and masks: 
\begin{enumerate}

\item The modulation of the power spectrum due to special relativistic effects caused by the motion of our Galaxy relative to the CMB rest frame, as proposed in \cite{Challinor:2002zh}, causes a small power asymmetry along the direction of motion, which was measured for the first time in \cite{Aghanim:2013suk}. We will refer to this effect as `relativistic power modulation'.

\item Sharp edges of the mask can have an effect on the power on small angular scales. If a mask has more sharp edges along one of the lines of sight being examined, or if the sharp edges of the mask happen to align with the anisotropies along one line of sight, this can cause the average power measured in a disc centered at that line of sight to be different. We will refer to this effect as the `edge effect'.

\item Although in a Gaussian, full sky, map, all multipoles are uncorrelated, our best estimate of the power spectrum at different scales will become weakly correlated if the full sky cannot be analysed, for example after a mask has been applied \cite{Wandelt:2000av}, or, more importantly, if the local power within a small part of the sky is analysed, as done here. Therefore a small amount of the large scale power asymmetry will get imprinted into the small scales. We will refer to this effect as `inter-scale correlations'.

\end{enumerate}

It is important to quantify the size of these three effects in order to learn about the amplitude of any intrinsic small scale asymmetry. We will discuss each effect in detail in the following section.

\subsection{Small scale asymmetry contaminants}

\subsubsection{Relativistic power modulation}

%%%%TAB%%%%%
\begin{table}[b]
\begin{center}
\begin{tabular}{ l  l  }
\toprule
$b_\smica$ & $\dbrel$\\ \midrule
2 & 0.43\% \\
3 & 0.65\% \\
4 & 0.86\% \\
5 & 1.08\% \\
\bottomrule
\end{tabular}
\caption{\label{table:dbrel} Contribution of the relativistic power modulation to the small scale asymmetry, $\dbrel$, for different values of the (unknown) boost factor in the \smica map, $b_{\smica}$.}
\end{center}
\end{table}
%%%%%%%%%%%%

The motion of our galaxy relative to the CMB rest frame does not only generate a dipole in the CMB. It also has two other special relativistic effects, an aberration effect and a power modulation effect. Those effects were first proposed in \cite{Challinor:2002zh}, and first measured by Planck \cite{Aghanim:2013suk}. The latter effect causes a power asymmetry in the CMB, which is strongest in the direction of the cosmological dipole, $(l,b)=(264\dg,48\dg)$. The temperature fluctuations are modulated by a factor 
\beq
{\delta T \sim (1+b_{\nu}\mathbf{n}\cdot\boldsymbol{\beta})},
\eeq
where $\mathbf{n}$ is the (normalised) direction of observation, $\boldsymbol{\beta}$ is the direction of the cosmological dipole (with an absolute value of ${\beta=1.23\cdot10^{-3}}$), and $b_{\nu}$ is a frequency dependent boost factor \cite{Aghanim:2013suk}.

The \smica map consists mainly of the 143\,GHz and 217\,GHz frequency bands \cite{Ade:2013hta}, for which $b_{143}\simeq2$ and $b_{217}\simeq3$, respectively. Higher frequency bands, for which the boost factor is much larger (as large as $\sim 10$ for 545\,GHz) are subdominant, but also present. It is therefore reasonable to estimate the boost factor of the \smica map as $2 < b_{\smica} < 5$. Unfortunately we cannot calculate the exact value because the multipole-dependent weights used for constructing the \smica map are not publicly available.

The power spectrum along a direction $\mathbf{n}$ is thus modulated in linear order (in $\beta$) by a factor $C_{\ell}^{\mathbf{n}}\sim(1+2b_{\nu}\beta\cos\chi)$, where $\chi$ is the angle between the direction of the cosmological dipole and the direction of observation. This modulation effect causes an asymmetry when comparing the power in two opposite directions in the CMB, which reduces in linear order to
\beq
\frac{C_{\ell}^{\mathbf{n}}-C_{\ell}^{-\mathbf{n}}}{C_{\ell}^{\mathbf{n}}} = 4b_{\nu}\beta\cos\chi + \mathcal{O}(\beta^2).
\eeq

Therefore, for a disc with $90\dg$ diameter centred at $(l,b)=(224\dg,0\dg)$ we expect a contribution to the asymmetry from the relativistic power modulation of $\dbrel = 4b_{\nu}\beta\langle\cos\chi\rangle$, where $\langle\cdot\rangle$ means the average over the disc. A straightforward calculation gives $\dbrel=0.43\%-1.08\%$, depending on the value $b_{\smica}=2-5$ (see Table \ref{table:dbrel}). We have to correct $\dbhl$ for this effect. Before we do that, we will also quantify the other contaminants of the small scale asymmetry.

Of course, the relativistic power modulation must also be taken into account when measuring the large scale asymmetry $\dbll$. Because the amplitude of the large scale asymmetry is quite high, this reduces the significances levels found with our measure only slightly ($2.8\sigma$ instead of $3\sigma$ for \smica mask and $2.4\sigma$ instead of $2.5\sigma$ for M74, for $b_{\smica}=2$).

\subsubsection{Edge effects}

\label{sec:edgeeffects}

%%%%Fig%%%%
\begin{figure}
\begin{center}
\includegraphics[scale=0.8]{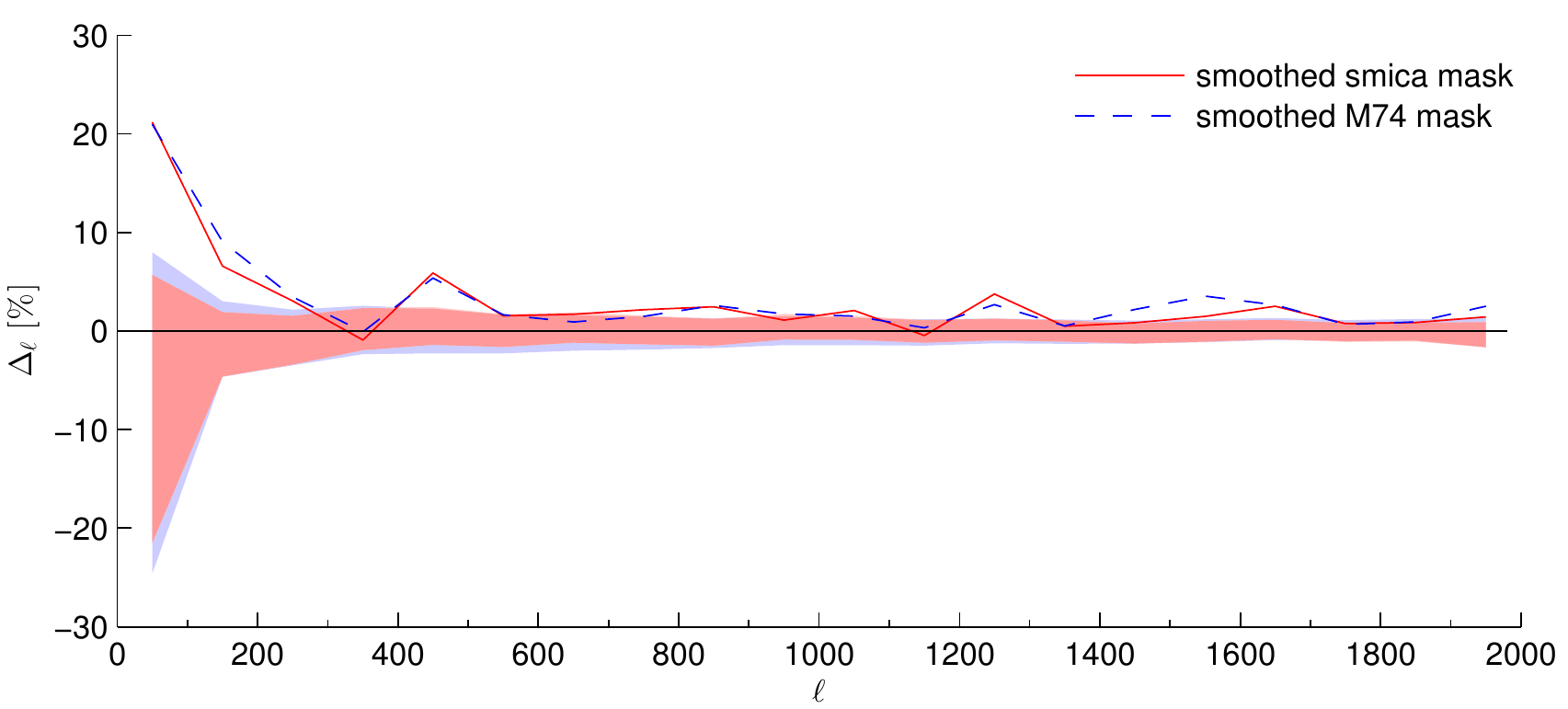}
\caption{\label{figure:Deltal_ssm_sM74} Binned plot of ${\Delta}_{\ell}$ along $(l,b)=(224\dg,0\dg)$ in the \smica map with smoothed \smica confidence mask (red, solid) and smoothed M74 mask (blue, dashed) applied. The red and blue bands are the $1\sigma$-regions from the corresponding values found in the simulations. This figure is to be compared to Fig.\,\ref{figure:Deltal_sm_M74}, where the masks have not been smoothed.}
\end{center}
\end{figure}
%%%%%%%%%%

%%%%Fig%%%%
\begin{figure}
\begin{center}
\includegraphics[scale=0.4]{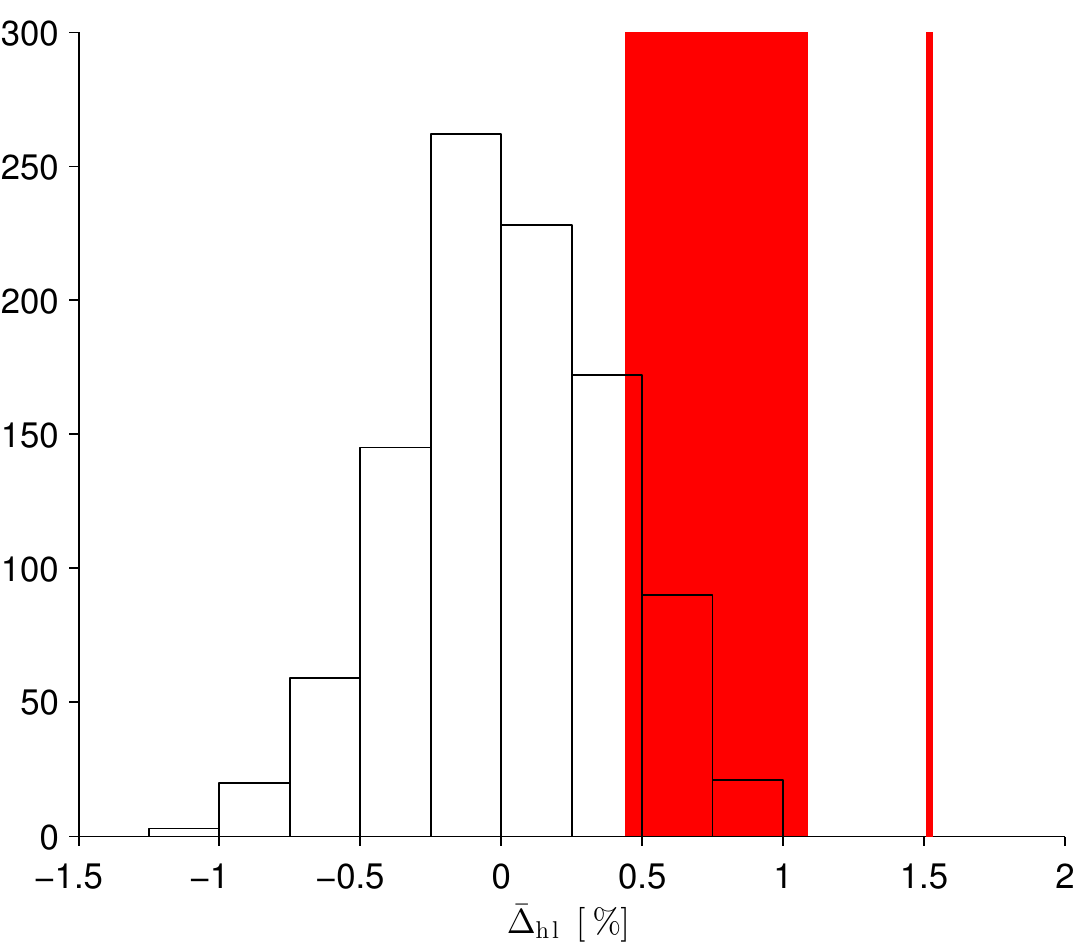}
\includegraphics[scale=0.4]{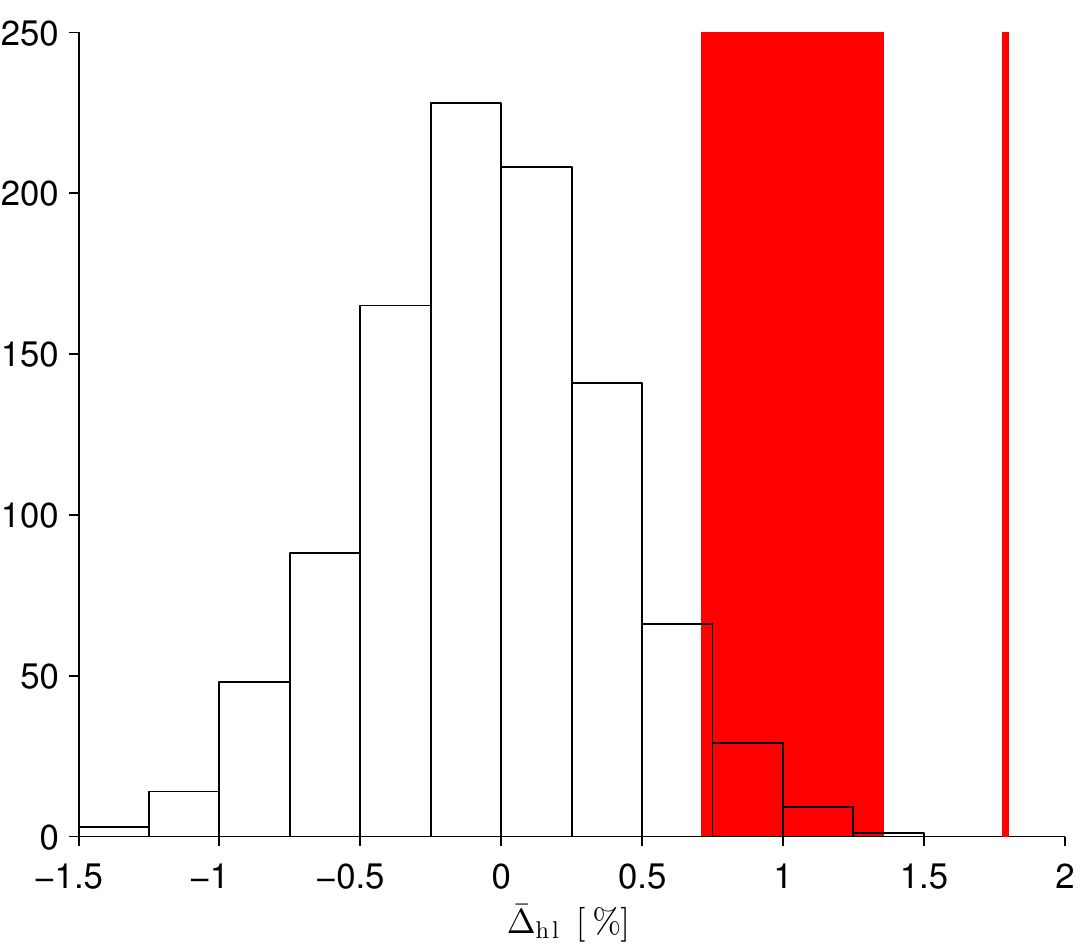}
\caption{\label{figure:hist_ssm_sM74} Distribution of $\dbhl$ in 1000 random CMB realisations with smoothed \smica mask (left) and smoothed M74 mask (right) applied. The red line indicates the value of $\dbhl$ in the \smica map, before correcting for the contribution from the relativistic power modulation. The red band indicates the value of $\dbhl$ after correcting for this effect, with the parameter $b_{\smica}=2$ on the right edge and $b_{\smica}=5$ on the left edge of the band. The true value lies somewhere in between.}
\end{center}
\end{figure}
%%%%%%%%%%

The mask used to block out galactic and point source foregrounds takes values of 1 or 0, which causes sharp edges in the cut sky map. These sharp edges might have an effect on the measurement of the power spectrum in high multipoles. If this is true, this might cause a power asymmetry between the discs used in our analysis, simply because in one disc there are more edges than in the other. This edge difference is evident in the \smica mask (see Fig.\,\ref{figure:locspots_smica}). In order to eliminate this possibility, we smooth the mask with a Gaussian filter before we apply it to the map. This way we remove all sharp edges in the cut sky map. We choose a FWHM of $10'$, which is larger than the pixel size ($\simeq1.72'$), and still small enough to keep possible foreground contamination negligible.

Using this method, we find  $\dbhl=1.52\%\,(4\sigma)$ with the smoothed \smica mask and $\dbhl=1.79\%\,(4.1\sigma)$ with the smoothed M74 mask, which corresponds to $p\sim10^{-5}$ (before correcting for the relativistic power modulation). The good agreement between the values for the two different smoothed masks indicates that all edge effects and thus the asymmetry caused by the mask have successfully been removed. This agreement is also reflected in the two plots of $\Delta_{\ell}$ as a function of ${\ell}$ (see Fig.\,\ref{figure:Deltal_ssm_sM74}): The two lines now lie almost on top of each other. Further, the agreement assures us that there is no strong foreground contamination caused by applying a smoothed mask. If there was such a contamination, it would be stronger when using the \smica mask, because it has a larger unmasked region than the M74 mask. Therefore we would observe a mismatch between the two values.

Note also that the monotonic increase of $\Delta_{\ell}$ with ${\ell}$, which made us suspicious in the first place, disappears after the masks have been smoothed, as can be seen in Fig.\,\ref{figure:Deltal_ssm_sM74}. This demonstrates how edge effects from the  mask boost the asymmetry along that line of sight at small angular scales.

We present histograms of the distributions for the smoothed masks in Fig.\,\ref{figure:hist_ssm_sM74}. These should be compared to the histograms in Fig.\,\ref{figure:hist_sm_M74}, where we did not smooth the masks (see also Table \ref{table:simulations} for statistical data from the simulations). It is evident that edge effects due to the mask cause a \emph{shift} of the mean (which is clearly visible in the case of the \smica mask), and a \emph{broadening} of the distribution. 

The \emph{shift} confirms what we proposed earlier and what is evident in Fig.\,\ref{figure:locspots_smica}, namely that the \smica mask simply has more edges in a disc centered at $(l,b)=(224\dg,0\dg)$ than in a disc centred at the opposite direction. Consequently, the values of $\dbhl$ in the \smica map, as well as in all of the random realisations, are shifted. Note that with the M74 mask applied, the mean of the simulations is actually slightly negative, meaning that the edge effects in this mask cause an asymmetry in the opposite direction. 

The \emph{broadening} on the other hand can be explained with the fact that different randomly realised maps can align differently with the sharp edges of the masks. This chance alignment is equally likely to happen along $(l,b)=(224\dg,0\dg)$ as it is along the opposite line of sight, leading to an increased probability of obtaining larger values of $|\dbhl|$ than we obtain for the smoothed mask. This is what is responsible for the increase in significance of the asymmetry found using the M74 mask when we smooth the mask. That is, the actual asymmetry in our universe remains; however the possibility of chance alignments in the random realisations mimicking this asymmetry is reduced. 

%%%%TAB%%%%%
\begin{table}%[!htbp]
\begin{center}
\begin{tabular}{ l c  c  c c c c }
\toprule
{}     	& \multicolumn{2}{c}{bare} & \multicolumn{2}{c}{sm} & \multicolumn{2}{c}{sm+con} \\ 
{}	    & $\mu$		&    $\sigma$   &   $\mu$ & $\sigma$      &    $\mu$  &  $\sigma$      \\
\midrule
\smica  & 2.24\% 	& 0.73\% 		& 0.02\%	& 0.37\% 		 & 0.46\% 	  & 0.36\% \\
M74     & -0.23\%	 & 1.22\% 		& -0.05\% 	& 0.45\%		& 0.65\%		& 0.42\% \\
\bottomrule
\end{tabular}
\caption{\label{table:simulations} Mean $\mu$ and standard deviation $\sigma$ of the small scale asymmetry $\dbhl$ in the 1000 random realisations of the CMB, with \smica mask (upper row) and M74 mask (lower row) applied. `bare' are the results without any corrections, `sm' are the results with smoothed masks, and `sm+con' are the results for the constrained realisations with smoothed masks.}
\end{center}
\end{table}
%%%%%%%%%%%%

Note however that there is a corresponding \emph{drop} in significance from the overwhelming value of $6.5\sigma$ to $4\sigma$ when using the smoothed \smica mask. This can only be explained by an extraordinary alignment between the edges in the \smica mask and the \smica map. It is worth stressing that this extraordinary alignment is extremely unlikely to occur by chance given the remarkably small $p$-value of $10^{-10}$ that corresponds to $6.5\sigma$. If it was simply due to chance, then one would expect some of our 1000 random realisations to also receive such chance alignments. This suggests that there is a systematic correlation between the \smica map and the edges of the \smica mask, introduced in the process of creation of the mask. 

This correlation might be explained as follows. The \smica mask is created from the \smica map by setting to zero the mask value in any pixel where the variance of the unmasked temperature map exceeds a certain threshold \citep{Ade:2013hta}. This way, Galactic and point source foregrounds are masked out. However, this also means that any masked region will be surrounded by a region of high variance in the temperature, slightly under the threshold. Edges in the \smica mask are thus correlated with regions of high variance in the \smica map. Such a correlation is not possible in any simulated map, and this is what might cause the significant boost of the small scale asymmetry in the \smica map. To fully examine this effect it would be necessary to add artificial foreground and simulate new masks for each of the 1000 random realisations using the same process used to create the \smica mask from the \smica map. It is beyond the scope of this work to perform such a test. It is satisfactory for our purposes that neither of the \emph{smoothed} masks show such a correlation between the observed CMB and the mask (as would be expected, because the smoothing process should wipe out any correlation on scales smaller than the smoothing scale of $10'$, which corresponds to ${\ell}=1080$). Our main results are presented using these smoothed masks.

Finally, note that edge effects affect only the small scales. We have checked that for the large scale asymmetry there is no significant change in the results if we use smoothed instead of unsmoothed masks. This is expected because the angular scales ($90\dg-0.3\dg$) corresponding to the low multipoles (${\ell}=2-600$) are larger than the smoothing scale ($10'\simeq0.17\dg$).

After removing the relativistic power modulation $\dbrel$ we find \emph{at most} (when using $b_{\smica}=2$) $\dbhl=1.09\%\,(2.9\sigma)$ with the smoothed \smica confidence mask and $\dbhl=1.36\%\,(3\sigma)$ with the smoothed M74 mask. This is still anomalous ($p\sim0.003$) and could be interpreted as independent evidence for a physical origin of the hemispherical asymmetry, if the small scale asymmetry was statistically independent from the large scale asymmetry. Unfortunately however, the large scales and small scales are \emph{not} completely independent, even in \lcdm, once a mask has been applied to the map \cite{Wandelt:2000av}, or when the local power inside a fraction of the sky is calculated. We need to correct for one more effect, the inter-scale correlations.

\subsubsection{Inter-scale correlations}

%%%%Fig%%%%
\begin{figure}
\begin{center}
\includegraphics[scale=0.6]{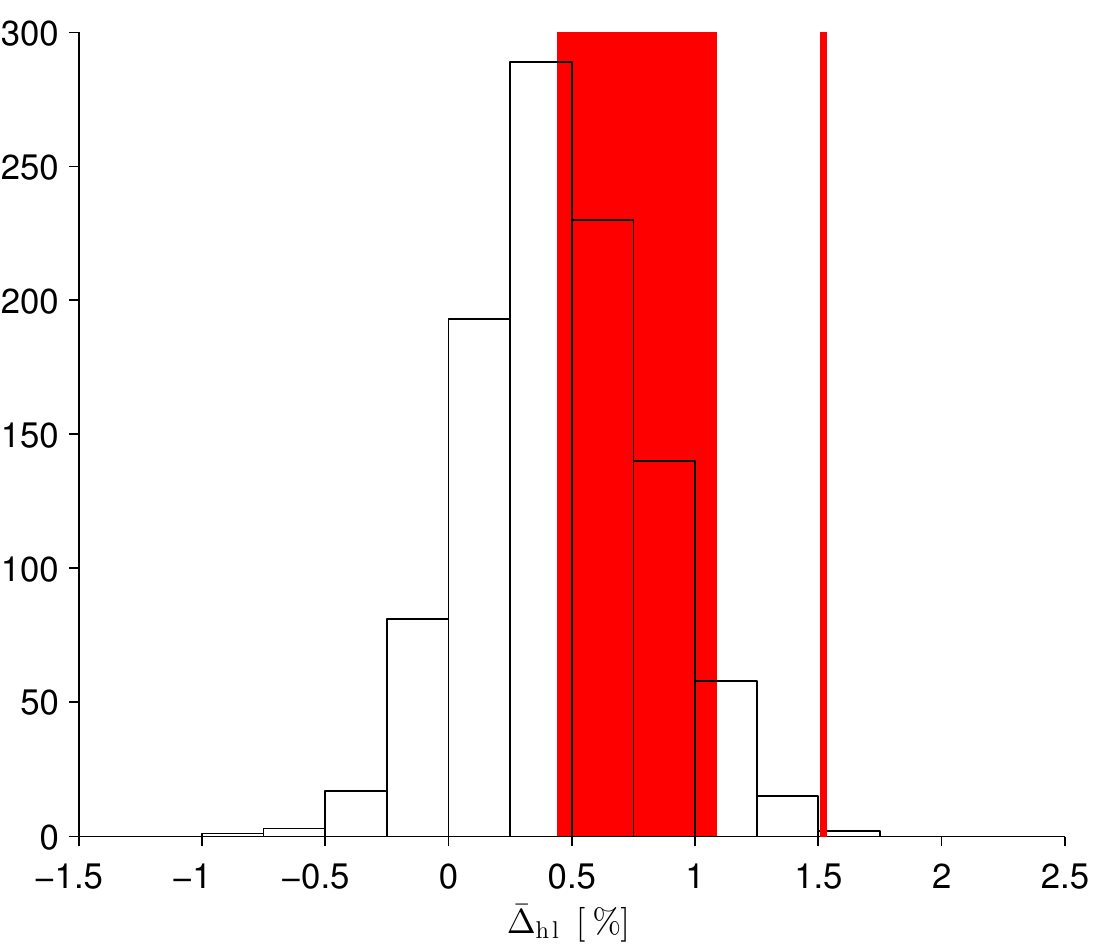}
\includegraphics[scale=0.6]{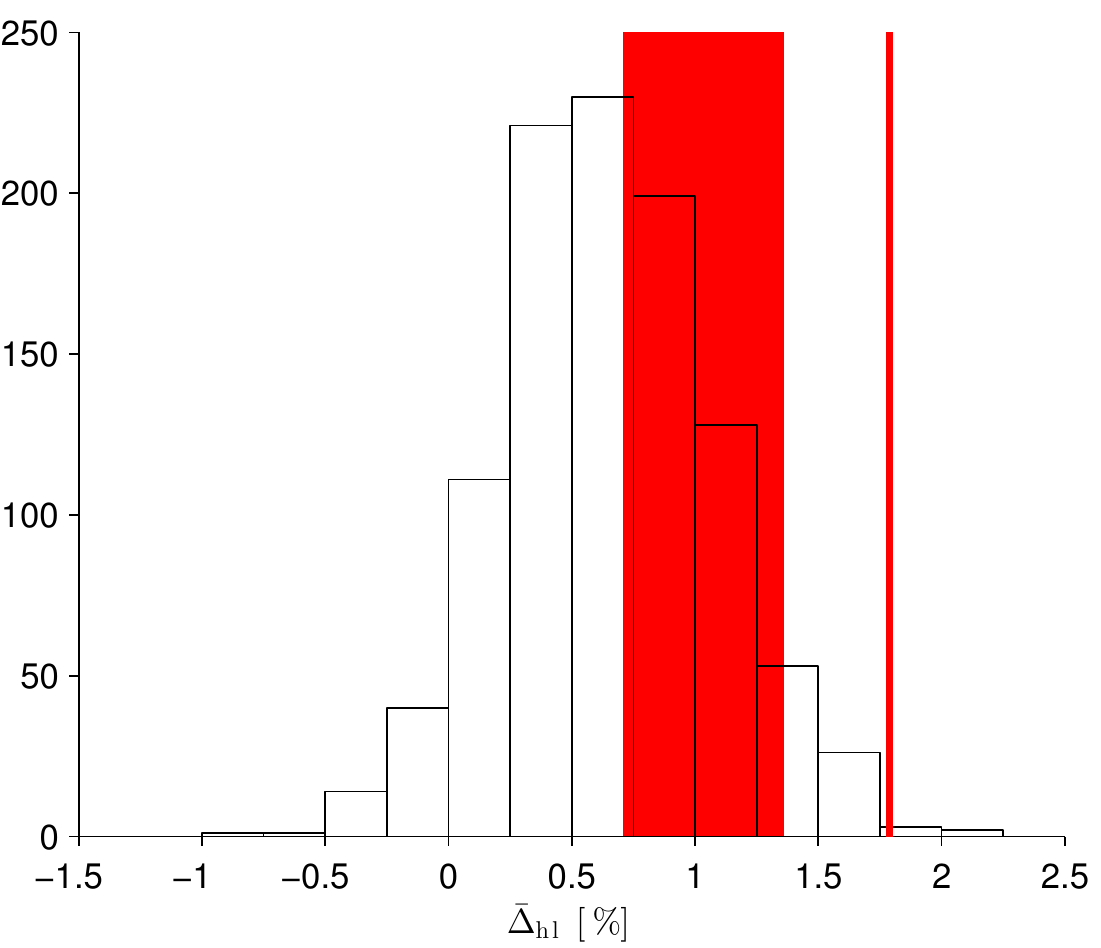}
\caption{\label{figure:hist_csm_M74} Distribution of $\dbhl$ values in 1000 constrained random CMB realisations with smoothed \smica mask (left) and smoothed M74 mask (right) applied. As before, the red line indicates the value of $\dbhl$ in the \smica map before correcting for the contribution from the relativistic power modulation, and the red band indicates the estimate of $\dbhl$ after correcting for this effect. The simulations were constrained to have a low scale asymmetry at least as strong as the one present in our CMB.}
\end{center}
\end{figure}
%%%%%%%%%%

%%%%TAB%%%%%
\begin{table}%[!htbp]
\begin{center}
\begin{tabular}{ l  c  c  c  c  c }
\toprule
{} & bare & sm & sm+rel(2) & sm+rel(2)+isc & sm+rel(3)+isc \\	 
\midrule
\smica  & 7.01\%\,($6.5\sigma$) & 1.52\%\,($4.0\sigma$) & 1.09\%\,($2.9\sigma$) & 0.63\%\,($1.8\sigma$) & 0.41\%\,($1.1\sigma$) \\ 
M74     & 4.24\%\,($3.6\sigma$) & 1.79\%\,($4.1\sigma$) & 1.36\%\,($3.1\sigma$) & 0.71\%\,($1.7\sigma$) & 0.49\%\,($1.2\sigma$) \\ 
\bottomrule
\end{tabular}
\caption{\label{table:summary} Summary of our results for the small scale power asymmetry $\dbhl$, using the \smica confidence mask (upper row) and the M74 mask (lower row). `bare' are the the results without any corrections. The other columns each add one correction:
`sm' - corrected for edge effects by applying smoothed masks.
`rel($x$)' - corrected for the relativistic power modulation by subtracting $\dbrel$, using $b_{\smica}=x$.
`isc' - corrected for the inter-scale correlations by subtracting $\dbisc$.}
\end{center}
\end{table}
%%%%%%%%%%%%

In the power spectrum corresponding to a full sky Gaussian map, the power in each of the multipoles is independent of the power in all of the other multipoles. However, in a cut sky map, e.g. after a mask is applied, large scale power gets weakly correlated with small scale power \cite{Wandelt:2000av}. This has also to be taken into account when calculating the localised power within a disc, as done here. Therefore a part of the excess small scale asymmetry observed in earlier sections might be induced by the excess large scale asymmetry. In this context, it would still be anomalous; however it would be a consequence of the anomalous large scale asymmetry.

In order to account for this possible effect, we have created 1000 \emph{constrained} random realisations of the CMB, which all have a large scale power asymmetry at least as large as the one present in the \smica map, i.e. realisations with the constraint that $\dbll \ge 6.9\%$. 

In order to reduce computational effort, we use a lower map resolution of $N_{\rm side}=256$ for calculating the low-$\ell$ asymmetry $\dbll$. If the simulation has a large enough low-$\ell$ asymmetry, we then calculate the high-$\ell$ asymmetry in the same map in full $N_{\rm side}=2048$ resolution. We downgrade the masks using the method proposed in \cite{Ade:2013nlj}: After downgrading from $N_{\rm side}=2048$ (pixel size $\simeq 1.72'$) to $N_{\rm side}=256$ (pixel size $\simeq 13.74'$) we set to zero every mask value smaller $0.8$ in order to avoid possible foreground contamination.

The mean of the $\dbhl$ values from the constrained realisations are clearly shifted to a positive value (see Fig.\,\ref{figure:hist_csm_M74}, also Table \ref{table:simulations}). Thus we confirm that large scale power gets correlated with small scale power in our analysis, so that a part of the small scale asymmetry is caused by the large scale asymmetry. We denote this contribution from inter-scale correlations as $\dbisc$. This number is given by the mean of the constrained simulations, i.e. $\dbisc=0.46\%$ for the smoothed \smica mask and $\dbisc=0.65\%$ for the smoothed M74 mask.

Comparing to the constrained realisations, the significance level of $\dbhl$ drops to $2.9\sigma$ for the \smica map with smoothed \smica mask and $2.7 \sigma$ for \smica map with smoothed M74 mask applied. After taking into account the relativistic power modulations, the significance level of $\dbhl$ drops to at most $1.8\sigma$ for the \smica mask and $1.7\sigma$ with the M74 mask. Note that this is the upper bound obtained for $b_{\smica}=2$. For $b_{\smica}=3$ the significance drops to $\sim1\sigma$. For detailed results see table \ref{table:summary}.

To summarise, we have shown that the apparent $6.5\sigma$ small scale asymmetry is in fact a coincidence of three effects, one physical effect (the relativistic power modulation) and two non-physical effects (edge effects and inter-scale correlations). The inter-scale correlations are a well-known effect arising from the study of angular harmonic functions on an incomplete sky \cite{Wandelt:2000av}; however the edge effects seem to imply a problematic correlation between sharp edges in the \smica mask and \smica map, which unnaturally enhances the power around the mask and on the smallest scales.

\section{Implications for theoretical models}

Theoretical models capable of explaining the hemispherical power asymmetry have arisen within the context of inflation. It has been shown in \cite{Erickcek:2008sm} that the hemispherical asymmetry cannot be generated during single-field slow-roll inflation without violating constraints on the homogeneity of the Universe. However, in the curvaton model such a power asymmetry can be produced without violating the homogeneity contraint. In particular, the asymmetry can arise due to a large-amplitude, superhorizon, perturbation in the curvaton field. The authors also show that this model makes other predictions regarding the CMB polarisation, which might be testable with future experiments. The analysis in \cite{Erickcek:2008sm} was repeated in \cite{Lyth:2013vha} with current data and weaker assumptions about the curvaton potential. It was shown that the induced power asymmetry can agree with the observation if $|\fnl|$ in the equilateral configuration is $\simeq 10$ on the Gpc-scale and $\lesssim 3$ on the Mpc-scale. A different model was recently proposed in \cite{Cai:2013gma}, where it was shown that the power asymmetry may be explained by a modulation of the sound speed of the inflation.

With our small scale null result we are able to put constraints on these models. Let, as before, $\dbrel$ be the contribution to the small scale asymmetry from the relativistic power modulation, and  $\dbisc$ be the contribution from inter-scale correlations. The intrinsic small scale asymmetry in the CMB is then given by $\dbhl-\dbrel-\dbisc$. Therefore, we can conclude, with a confidence level corresponding to a statistical significance of $n\sigma$, that any small scale asymmetry generated by any theoretical model must be below $\dbhl-\dbrel-\dbisc+n\sigma$ (where $\sigma$ in this context comes from the \emph{constrained} random realisations). In particular, with our measure we restrict the power asymmetry in the high multipoles ${\ell}=601-2048$ to be smaller than $1.53\%$ at $95\%$ C.L. This value was derived using $b_{\smica}=2$ and the M74 mask applied, which gives the largest and therefore most conservative upper bound at this level of confidence.

The hemispherical asymmetry is typically described as a dipolar modulation of the form $\delta T \sim (1+A(k)\mathbf{p}\cdot\mathbf{n})$, where $\mathbf{p}$ is the (normalised) direction of asymmetry, $\mathbf{n}$ is the (normalised) direction of observation and $A(k)$ is a free parameter of the model, which describes the amplitude of the modulation and can depend on the scale. We already discussed the effect of such a modulation on our asymmetry measure earlier, in the context of the relativistic power modulation. Following the same derivation, we have $\dbhl=4A(k)\langle\cos\chi\rangle$, where $\chi$ is the angle between the direction of observation $\mathbf{n}$ and the direction of the asymmetry $\mathbf{p}$ (here given by $(l,b)=(224\dg,0\dg)$), and $\langle\cdot\rangle$ means the average over the disc of diameter $90\dg$ centered at $(l,b)=(224\dg,0\dg)$. In order to translate multipoles into scales, we use the relation $\ell\simeq x_{\rm ls} k$, where $x_{\rm ls}\simeq14\,\rm Gpc$ is the distance to the last scattering surface. Using these translations, we find that, on average, $A(k)<0.0045$ in the multipoles ${\ell}=601-2048$, which corresponds to scales $7-23\,\rm Mpc$, at $95\%$ C.L. This new constraint must be followed by any model aiming to explain the origin of the large scale hemispherical asymmetry. Note that our constraint on $A(k)$ is even tighter than the constraint from the distribution of distant quasars, $|A(k)|<0.012$ on the Mpc-scale ($95\%$ C.L.) \cite{Hirata:2009ar}. 

For completeness we also observe that in the lowest multipoles, ${\ell}=2-100$, we determine an average asymmetry of $\sim20\%$ (see Fig.\,\ref{figure:Deltal_sm_M74}), which translates into $A(k)\sim0.06$. This number is consistent with the number given in the recent Planck results, $A=0.07\pm0.02$ for ${\ell}<{\ell}_{\rm max}=64$ \cite{Ade:2013nlj}. The amplitude of the dipolar modulation is thus strongly scale-dependent; it must go to zero at small scales (high multipoles). Such a scale-dependence of the power modulation (or lack of asymmetry on small scales) is also found by analysing the trispectrum of the CMB from the 143\,Ghz and 217\,Ghz frequency maps directly, see Ref.\,\cite{Ade:2013ydc}, especially Fig.\,16 therein.

%----------------------------------%
\section{Discussion}

Statistical anomalies that arise in cosmological observables are certainly interesting. This is primarily because they might be hints of new physics beyond \lcdm. However, we need to keep in mind that in a large cosmology community statistical flukes can and will be found. Therefore it is of prime importance to look at existing anomalies in as many different ways as possible to obtain new information about them.

Our motivation in this work has been to obtain more information about the large scale hemispherical power asymmetry by studying the power on small scales at the same line of sight. We introduced a simple measure of the overall asymmetry, the normalised difference between the average power in two opposite discs in a map of the CMB. Using this measure, we confirm the existence of a large scale hemispherical asymmetry with a statistical significance of $\sim3\sigma$ ($p \simeq 0.003$) in the multipole range ${\ell}=2-600$.

Applying the same measure to the high multipoles ${\ell}=601-2048$, we found an anomalously high asymmetry, with a naive statistical significance as large as $6.5\sigma$ ($p\sim10^{-10}$). We showed however that this extreme anomaly is a coincidence of three different effects: relativistic power modulation, edge effects caused by the mask, and inter-scale correlations, which naturally occur when analysing an incomplete sky. After correcting for all of these effects, the significance of the anomaly drops to $\sim 1\sigma$, in other words, there is no intrinsic anomaly in the small scales (high multipoles).

It is worth pointing out that the initial apparent $6.5\sigma$ anomaly can only be explained by an extraordinary alignment between the \smica map and the edges of the \smica mask. In fact, after removing the sharp edges from the mask, the significance drops to $\sim4\sigma$. This indicates the existence of an unphysical correlation between the \smica map and its confidence mask. 

In \cite{Ade:2013nlj}, the Planck collaboration calculated the dipole directions for 100-multipole bins of the local power spectrum distribution up to ${\ell}=1500$. Even for the highest of those multipole bins, the dipole directions seemed to cluster around $(l,b)=(224\dg,0\dg)$ (see Fig.\,27 therein). This is what we would expect from our `bare' results, where the asymmetry in the high multipoles is very strong along that direction, due to the effects discussed. The similarity to our bare results is also clear by comparing our Fig.\,\ref{figure:Deltal_sm_M74} to Fig.\,28 in \cite{Ade:2013nlj}. Thus it seems to us that the corrections discussed here were not taken into account in \cite{Ade:2013nlj}. In fact, according to our results such a clustering direction should be around $(l,b)=(224\dg,0\dg)$ at the low multipoles (${\ell}<600$), where the hemispherical asymmetry is dominating, and $(l,b)=(264\dg,48\dg)$ at the high multipoles (${\ell}>600$), where the relativistic power modulation is dominating. It is interesting to note that in \cite{Aghanim:2013suk}, where a similar analysis was performed, exactly this behaviour was found (see Fig.\,3 therein).

Our result has important consequences for theories. Any theory aiming to explain the large scale hemispherical asymmetry must not produce a small scale asymmetry larger than our constraint (with our measure), $1.53\%$ at $95\%$ C.L., in the high multipoles ${\ell}=601-2048$. This can be translated into a constraint for the modulation amplitude $A(k)$ in models which fit the hemispherical asymmetry to a dipolar modulation, giving $A(k)<0.0045$ on $\sim$10\,Mpc-scales, at $95\%$ C.L. 

The full impact of our result on the theoretical side remains to be investigated. However we note that the theoretical models that can generate the large scale asymmetry typically also require a large value for the non-Gaussianity parameter $\fnl$. Our results would therefore require a strong scale-dependence in this parameter so as that the large scale asymmetry does not propagate into the small scales. It is curious to note that the constraints on $\fnl$ from analysis of the CMB bispectrum also seem to favour a larger $\fnl$ value at larger scales (e.g. see figure 20 in \cite{Ade:2013ydc}).
\vspace{1cm}

\subsection*{Additional remark}
We also analysed, using the same method discussed, the power asymmetry in Planck's lensing map, which maps the distribution of all matter between the last scattering surface and us. As proposed in \cite{Ade:2013tyw}, we reduced the (dominating) noise in that map by applying a Wiener filter given by the theoretical expectation of the angular power spectrum of the lensing potential. We did \emph{not} find a significant asymmetry along $(l,b)=(224\dg,0\dg)$, compared to other lines of sight within that map. This suggests that there is no preferred direction in the distribution of large scale structure along that line of sight which could cause the hemispherical asymmetry in the CMB.

%-------------Acknowledgments------------------% 
\acknowledgments
We thank Dominik Schwarz and Seshadri Nadathur for helpful discussions and comments, and Duncan Hanson for help generating the Planck lensing map. SF acknowledges support from the doctoral programme in particle- and nuclear physics (PANU). SH acknowledges support from Academy of Finland grant 131454.  Some of the results in this paper have been derived using the \healpix software package (K.M. Gorski et al., 2005, ApJ, 622, p759).

\bibliography{asym.bib}
\bibliographystyle{JHEP}

\end{document}